\let\origvec\vec
\documentclass[conference]{IEEEtran}

\let\vec\origvec
     \usepackage{amssymb,amsmath}
\usepackage[T1]{fontenc}
 \usepackage{graphicx}
\usepackage{mdwmath}
\usepackage{mdwtab}
\usepackage{soul}
\usepackage{cite}
 \usepackage{multirow}
 \usepackage{color} 
\begin{document}
\title{Mitigating Censorship with Multi-Circuit Tor and Linear Network Coding}
%\author{Anna Engelmann\orcidID{0000-0001-6806-0155} \and
%Admela Jukan\orcidID{0000-0002-4434-6340}}
%%
%\authorrunning{A. Engelmann et al.}
%% First names are abbreviated in the running head.
%% If there are more than two authors, 'et al.' is used.
%%
%\institute{Technische Universit\"at Carolo-Wilhelmina zu
%Braunschweig\\ Hans-Sommer-Str. 66, 38106 Braunschweig, Germany\\
%\email{\{a.engelmann, a.jukan\}@tu-bs.de}}
%%
%\maketitle              % typeset the header of the contribution
%%

\author{\IEEEauthorblockN{Anna Engelmann and Admela Jukan}
\IEEEauthorblockA{Technische Universit\"at Carolo-Wilhelmina zu
Braunschweig, Germany\\
Email: \{a.engelmann, a.jukan\}@tu-bs.de}
}
\maketitle

\begin{abstract}
Anonymity networks are providing practical mechanisms to protect its users against censorship by hiding their identity and information content. The best-known anonymity network, The Onion Routing (Tor) network, is however subject to censorship attacks by blocking the public Tor entry routers and a few secret Tor entry points (bridges), thus preventing users to access the Tor. To further advance the evolution of anonymity networks, while addressing censorship attacks, we propose to enhance the well-known multi-circuit Tor technique with linear network coding (LNC) and analyze the resulting censorship success. The results show that LNC can improve the robustness of Tor against censorship.\\
%{\keywords{Anonymity, Censorship prevention, The Onion Routing.}}
\end{abstract}

\section{Introduction}
\par The Onion Routing (Tor) provides practical mechanisms to provide user anonymity and thus mitigate censorship. In Tor, traffic tunneling is provided over a chain of selected onion routers (OR), which in combination with layered encryption prevent linking tof communicating parts or decrypting the information exchanged \cite{Erdin:2015}. Thus, Tor is designed to alleviate censorship by hiding user's identity and the content communicated. On the other hand, censorship in Tor is de-facto possible to implement,  the most known example of which is "Great Firewall of China," where a few major points in the network are deployed to filter and block the incoming and outgoing Internet traffic,  thus preventing users to connect to Tor. 

To mitigate censorship, Tor has introduced bridges, in form of non-public onion routers that are known to censored users only and can still allow them an entry into the Tor network \cite{AlSabah:2013}. Nevertheless, a censoring entity can collect the said bridges and block them as well. To decrease the chances of blocking, while improving Tor performance and randomizing the tunnel (circuit) distribution, multipathing in Tor has been proposed \cite{Karaoglu:2012,Rochet:2015,Yang:P:2015, Yang:2015}. Even with multipathing, however, there is no mechanism as of today to recovering if any of the blocked circuits or lost traffic in case of censorship. 

\par In this paper, we hope to further contribute to Tor evolution by improving the defying censorship with multipath routing, with an addition of Linear Network Coding (LNC). In our approach, LNC is used in combination with multi-circuit Tor to recover the traffic blocked by censorship. To evaluate the benefits of this idea, we define and compare three Tor implementations: 1) one Tor (oTor), i.e., the traditional Tor implementation with one communication circuit; 2) multi-circuit Tor (mTor), where traffic transmission is implemented over multiple Tor circuits and diverse ORs; 3) coded Tor (cTor), where mTor traffic is encoded with LNC before transmission. The results show that LNC in cTor can more effectively mitigate censorship as compared to Tor and mTor.

\section{Multiple-Circuit Tor with LNC (cTor)}
\begin{figure*} [t]
  \vspace{-0.2cm}\centerline{\includegraphics[width=0.7\textwidth]{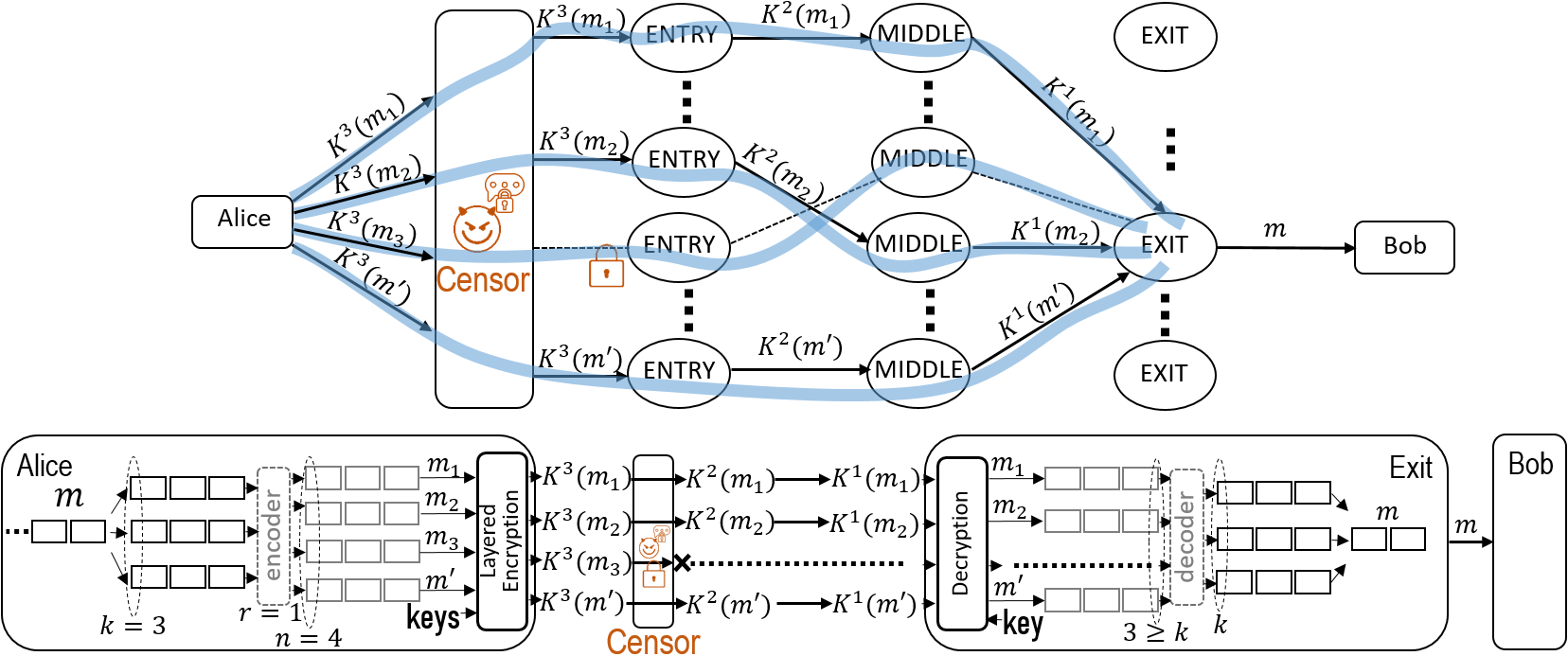}}
  \vspace{-0.2cm}
  \caption{\small Reference architecture of Tor with multiple circuits and LNC, i.e., cTor.}
  \label{cTor1}\vspace{-0.5cm}
  \end{figure*}

Fig. \ref{cTor1} illustrates our proposal to implementing LNC in mTor network (cTor). Just like in oTor, -- the basic Tor architecture, also mTor and cTor include Onion Proxies (OP) and Onion Routers (OR). Let us assume that OP includes a client (Alice), who initiates the anonymous communication and uses information about existing ORs to setup circuits to server (Bob) over three randomly selected ORs (entry or bridge, middle and exit ORs). The traffic is split into fixed-sized (512-byte) units called \emph{cells}, which are encrypted in layers with keys of entry, middle and exit ORs so that each OR can remove only one encryption layer applying the same key as Alice. In contrast to Tor, where all packets are sent over the same circuit, a client in mTor and cTor splits incoming traffic $m$ into cells, builds $n$ traffic sub-flows and sends them through $n$ circuits toward one exit node. Alice randomly selects $2n+1=9$ ORs and setups $n=4$ disjoint circuits, which only share a common exit router. The third circuit contains bridge known to the censor and will be blocked resulting in cell loss. 

\par The lower part of Fig. \ref{cTor1} shows how Linear Network Coding (LNC) can be used in the system as erasure code and can be utilized to protect traffic against losses. Generally, any ($n$, $k$) erasure code encodes $k$ units of original data into $n$ units of coded data and tolerate lost of up to $r=n-k$ data units. Thus, Alice splits incoming traffic $m$ into cells and parallelizes them to build $k=3$ cell sub-flows, e.g., $m_1, m_2,m_3$. The encoder takes one cell from each sub-flow and encodes $k$ cells to generate $r=1$ redundant cells, i.e., sub-flow $m'$. We refer to any $k$ cells of original data encoded together as \emph{generation}. After encoding, $n=k+r$ coded cells leave the encoder building $n=4$ parallel sub-flows. Each coded sub-flow is finally encrypted in layers to be sent over certain circuit toward exit OR. Due to censorship, the third circuit is interrupted resulting in loss of $m_3$. Circuit failures and cell losses, however, do not have any impact on the throughput of cTor as long as at least $k=3$ circuits are able to deliver coded cells to exit node. During decoding process any $k=3$, e.g., $m_1, m_2, m'$, out of $n=4$ cells from the same generation can recover the original cells. Finally, the recovered cells can be serialized into original $m$ and sent to Bob. In contrast, the circuit blocking in Tor and mTor will result in communication interruption, since the exit node will be unable to recover original $m$ due to missing cells and has to request retransmission. 

\section{Analysis of censorship success and evaluations }
\par We assume that there are $M_b$ unknown bridges and $M'_b$ bridges known to a censor. The censored user randomly selects $n$ entry nodes out of $M_b+M'_b$ bridges resulting in $\binom{M_b+M'_b}{n}$ possible bridge combinations. The censorship is successful, if the anonymous communication was successfully disrupted by bridge blocking, i.e., the receiver is not able to recover the sent information. Blocking of entry ORs, i.e., bridges, results in blocking of the circuit and loss of whole cell flow sent over this circuit. In case of the anonymous communication over Tor and mTor, the communication between client and server is disturbed, if at least one bridge utilized as entry OR is known to a censor and blocked. The client randomly selects $n$, where $n\geq 1$, out of $M_b+M'_b$ available bridges, whereby the probability to select known to the censor and blocked bridge, i.e., the probability for a censorship success, can be calculated as
%\begin{equation}\label{Cens}% Eq(1) TESTED
$P_{bb}=\tfrac{1}{\binom{M_b+M'_b}{n}}\sum_{i=1}^{min\{n, M'_b\}}\binom{M_b}{n-i}\binom{M'_b}{i}$. Thus, client selects $n-i$ honest and $i$ censored bridges out of $M_b$ and $M'_b$, respectively.
When LNC is applied (cTor), the censorship is only successful if more than $r$ out of $n$ utilized bridges are known to the censor and blocked. Thus, the destination receives less than $k$ cells from each generation and can not recover the original information by decoding, i.e., the probability for successful censorship is 
%\begin{equation}\label{CensC}%TESTED
$P^{\text{LNC}}_{bb}=\frac{1}{\binom{M_b+M'_b}{n}}\sum_{i=r+1}^{min\{n, M'_b\}}\binom{M_b}{n-i}\binom{M'_b}{i}$, $\text{   }M'_b>r$.
%\end{equation}
%, whereby the cTor communication can not be disrupted by bridge blocking when $M'_b\leq r$.

%\section{Performance evaluation}
\par We now analyze a generic network topology, whereby any client can select between $M_b+M'_b\geq25$ bridges. %Since all simulation results obtained with 95\% confidence interval conform to the analytical results, they are not shown here for clarity of illustration. 
   \begin{figure} [t]
  \centerline{\includegraphics[width=0.95\columnwidth]{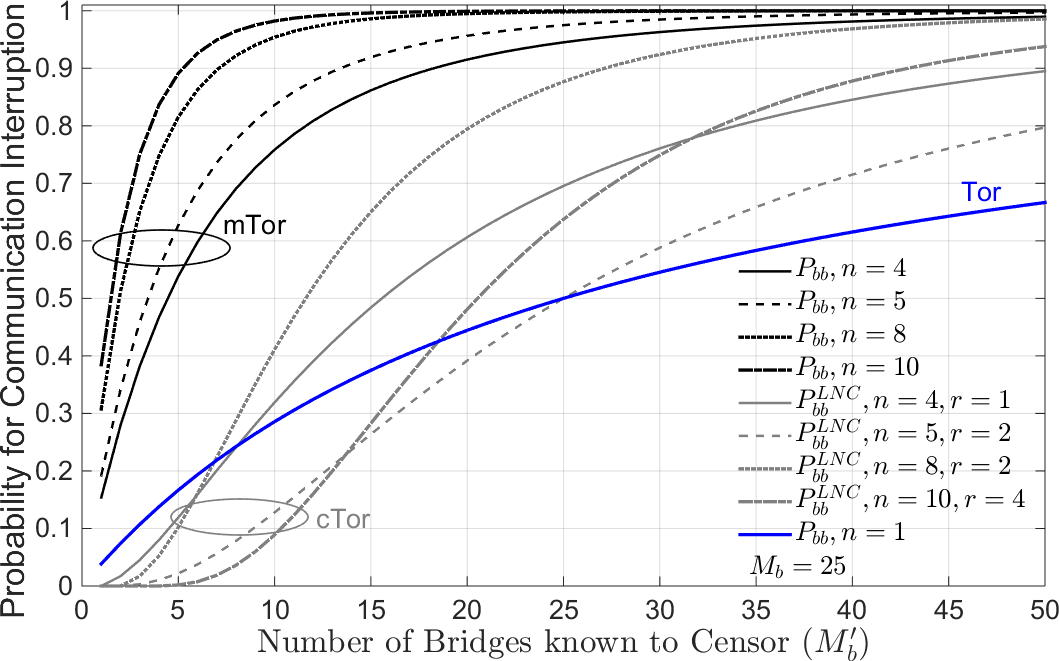}}\vspace{-0.2cm}
  \caption{\small Probability of successful communication interruption ($P_{bb}$ and $P^\text{LNC}_{bb}$).}
  \label{blocking1}\vspace{-0.5cm}
 \end{figure}
%%     \begin{figure} [t]
%  \centerline{\includegraphics[width=0.5\columnwidth]{fig/res/JammingCensor}}\vspace{-0.2cm}
%  \caption{\small Probability of successful communication interruption ($P_{jam}$ and $P^\text{LNC}_{jam}$) due to jamming of Tor traffic on censored circuits.}
%  \label{jamming}\vspace{-0.5cm}
%  \end{figure}
%  \begin{figure}
%    \subfigure[a]{\includegraphics[width=0.49\textwidth]{fig/res/blockingCensor}}
%    \subfigure[b]{\includegraphics[width=0.49\textwidth]{fig/res/JammingCensor}}
%\caption{Titel unterm gesamten Bild}
%\end{figure}
Fig. \ref{blocking1} shows the probability for successful communication interruption due to blocking of entry bridges, whereby we assumed that $M_b=25$ bridges are unknown to the censor. The probability of circuit blocking increases with increasing number of known bridges and utilized circuits $n$ related to Tor and mTor. The mTor communication over $n=8, 10$ circuits will be blocked with probability 100\% when the censor knows more than 15 bridges. In contrast, cTor shows much better performance, which depends of the amount of utilized coding redundancy $r$, e.g., cTor always outperforms mTor and Tor if $M'_b\leq 25$. The communication over $n=10$ and $n=5$ circuits with $r=4$ and $r=2$ redundancy shows the lowest probability of communication interruption by censorship as long as $M'_b<15$ and $15\leq M'_b<25$, respectively. 

\section{Conclusion}
We investigated multi-circuit Tor in combination with LNC to increase robustness against censorship. The results showed that cTor with random selection of ORs and circuits carries potential to significantly improving the robustness of anonymous communication against censorship as compared to Tor and mTor.

%\bibliographystyle{IEEEtran}
%\bibliography{codingbibTran}
\end{document}